\documentstyle{article}
\font\bb=msym10
\newcommand{\cn}{\mbox{\bb C}}
\newcommand{\rn}{\mbox{\bb R}}

\newtheorem{thm}{Theorem}[section]
\newtheorem{defi}[thm]{Definition}
\newtheorem{lemma}[thm]{Lemma}

\newtheorem{prop}[thm]{Proposition}
\newtheorem{cor}[thm]{Corollary}
\begin{document}
\bibliographystyle{plain}
\setlength{\baselineskip}{7 mm}

\title{Wave equations on q-Minkowski space}
\author{Ulrich Meyer \\
 {University of Cambridge}\\
{Department of Applied Mathematics  and Theoretical Physics} \\
{Cambridge CB3 9EW, England} \\
{\small e-mail U.Meyer@amtp.cam.ac.uk}}
\date{March 1994}
\maketitle
\vspace{-8cm} \hfill DAMTP/94-10 \vspace{8cm}

\begin{abstract}
We give a systematic account of the exterior algebra of forms on q-Minkowski
space,
introducing the q-exterior derivative, q-Hodge star operator, q-coderivative,
q-Laplace-Beltrami operator and the q-Lie-derivative. With these tools at hand,
we then
 give a detailed exposition of the q-d'Alembert and q-Maxwell equation. For
both
equations we present a q-momentum-indexed family of plane wave solutions.
We also discuss the gauge freedom of the q-Maxwell field and give a  q-spinor
analysis
of the q-field strength tensor.
\end{abstract}

\section{Introduction}\label{prelim}

In a previous paper \cite{DAMTP/93-45}, we gave a detailed account of the
q-deformation of spacetime and its symmetry group (see \cite{DAMTP/93-68} for a
detailed comparison with the approach of
\cite{Schlieker/3/90,Schlieker/6/91,Wess/92}).
The physical motivation
behind this  was to develop a q-regularisation scheme and/or give
a toy model for Planck scale corrections to the geometry of spacetime. As a
next step
we now investigate wave equations on this non-commutative spacetime.

The key idea in \cite{DAMTP/92-48,DAMTP/93-45} was that
q-Minkowski space should be given by $2\times 2$ braided Hermitean matrices,
which
were introduced by S. Majid in \cite{Majid/12/91} as a non-commutative
deformation
of the algebra of complex-valued polynomial functions on the space of ordinary
Hermitean matrices, i.e. on Minkowski space. Braided matrices have a central
and
grouplike element, the so-called {\it braided determinant}, which plays the
r\^{o}le of a q-norm and which determines a q-deformed Minkowski metric.
 One considers braided matrices and not
the  algebra of  $2\times 2$ {\it quantum} matrices because of
 the insufficient covariance properties of the latter \cite{DAMTP/92-12}.

However,   the braided matrices as given in
\cite{Majid/12/91}  did not generalise the additive group structure of
Minkowski
space. The addition of matrices should be reflected in our dual and q-deformed
setting
by a {\it braided coaddition} as introduced in \cite{DAMTP/92-65}, by which one
means
 a braided coproduct of the form
$\underline{\Delta} x = x \underline{\otimes} 1+1\underline{\otimes} x$ which
extends as an algebra map with respect
to a {\it braided tensor product} $\underline{\otimes}$ and not the ordinary
tensor product
$\otimes$. A braided tensor product $\underline{\otimes}$ is like the super
tensor products
encountered in the theory of superspaces, but with the $\pm 1$ factors replaced
by braid statistics. Explicitly, there is a {\it braiding } $\Psi$ such that
$$(a\underline{\otimes} b)(c\underline{\otimes} d) = a\Psi
(b\underline{\otimes} c)d,$$
i.e.  $\Psi$ measures how two independent copies of a system fail to commute.
In the
 case of the commutative algebra of polynomial functions on a space,
 $\Psi$ is simply given by the twist map $\Psi (a\otimes b )=b\otimes a$.
This braiding is determined by a background quantum group, which acts
as the symmetry group of the system. See \cite{Majid/6/92} and the references
therein
for an introduction to the theory of braided matrices and braided groups.

In \cite{DAMTP/93-45} we found  such a braiding
and background quantum group, which allowed for quantum Minkowski space to have
a braided coaddition. This gave rise to a natural quantum Lorentz group which
preserves the entire structure of quantum Minkowski space, i.e. both its
braided
coaddition and its non-commutative algebra structure. The final result is given
in
terms of two solutions of the 4-dimensional QYBE:
$$
\begin{array}{rcl}
{\bf R}^{\;\; ab}_{ M\; cd} & =& {\bf R}^{-1 L C}_{\;\;\;\;\;BI}
{\bf R}^{B^{'}I}_{\;J A}{\bf R}^{A^{'}J}_{\;KD^{'}}
\widetilde{{\bf R}}^{K D}_{\;C^{'}L},\\
 {\bf R}^{\;\;ab}_{L\;\;cd}&=&
{\bf R}^{C I}_{\;JB}{\bf R}^{B^{'}J }_{\;K A}
{\bf R}^{A^{'}K }_{\;L D^{'}}\widetilde{{\bf R}}^{L D}_{\;C^{'}I },
\end{array}
$$
Here $P$ denotes the permutation map and
\begin{equation}\label{rmat}
R =  \left( \begin{array}{cccc} q &0 &0 &0  \\0 & 1
&q-q^{-1} & 0\\
0 &0& 1& 0\\0&0&0&q\end{array}\right),\;\;\;\;\;q\in\rn
\end{equation}
is the standard $SU_{q}(2)$ R-matrix. The matrix $\widetilde{{\bf R}} $ is
defined as
$(({\bf R}^{t_{2}})^{-1})^{t_{2}}$, where $t_{2}$ denotes transposition
in the second tensor component.
We also used multi-indices $a = (A  A^{'}) = (11),(12),(21),(22)$.
These two matrices obey the   relation
\begin{equation}\label{quasiheck}
0=(P{\bf R}_{L}+1)(P{\bf R}_{M} -1),
\end{equation}
which  ensures the existence of a braided coaddition  \cite{DAMTP/92-65}.
In terms of these data, q-Minkowski space $M_{q}$ is given as the algebra of
quantum
covectors
$M_{q}=V^{'}({\bf R}_{M})$
 in the notation of \cite{DAMTP/92-12}. It has generators $x_{a}$ and a star
structure
$\bar{x}_{a}=x_{\bar{a}}$, where $\bar{a}= (A^{'}A)$ denotes the twisted
multi-index.
We denote the star structure by a bar in order to avoid confusion with the
Hodge
star operator.

The quantum Lorentz group ${\cal L}_{q}$ is defined as a quotient of the FRT
algebra $A({\bf R}_{L})$ with generators $\lambda^{a}_{\;b}$
by the metric relation $\lambda^{a}_{\;c}\lambda^{b}_{\;d}g^{cd}=g^{ab}$,
where the q-metric $g^{ab}$ is given by
$$
g^{ab} = \frac{1}{q+q^{-1}}\varepsilon_{AC}
R^{ A^{'}C}_{\; D B}
\varepsilon^{D B^{'}}
$$
in terms of the $SL_{q}(2,\cn)$-spinor metric
\begin{equation}\label{spim}
\varepsilon_{AB}=
\left( \begin{array}{cc}0&1/\sqrt{q}\\-\sqrt{q}&0 \end{array}\right).
\end{equation}
We are working in a `spinorial basis', where the metric has two negative and
two
positive eigenvalues. In this basis, the injective
$\ast$-Hopf algebra  morphism \cite{DAMTP/93-45} (which induces a push forward
of ${\cal L}_{q}$-comodules)
\begin{equation}\label{pushforward}
{\cal L}_{q} \hookrightarrow SL_{q}(2,\cn)
\end{equation}
has the simple form
 $\lambda^{a}_{\;b}\mapsto t^{\dagger B}_{\;\;\;A}t^{B^{'}}_{\;A^{'}}$,
where $t$ are the generators of $SL_{q}(2,\cn)$.
One can also choose a `$x,y,z,t$-basis' by a simple change of generators
\cite{DAMTP/93-45},
but for our purposes it is more convenient to stay in the spinorial basis.

If we now consider the coaction of ${\cal L}_{q}$ on $M_{q}$, one problem
arises:
In order to obtain full covariance under the coaction by the q-Lorentz group,
we have to adjoin ${\cal L}_{q}$ slightly by a single invertible central and
grouplike
element $\varsigma$ \cite{DAMTP/92-65}. The extended
q-Lorentz group is denoted by $\widetilde{\cal L}_{q}$, and its covariant right
coaction on q-Minkowski space is given by
$$\beta_{M_{q}} : x_{a}\mapsto x_{b}\otimes\lambda^{b}_{\;a}\varsigma .$$
Equation (\ref{pushforward}) implies that we also have a covariant coaction by
the extended algebra $\widetilde{SL}_{q}(2,\cn)$ given by
$x^{A}_{\;A^{'}} \mapsto
 x^{B}_{\; B^{'}} \otimes t^{\dagger A}_{\;\;\;B}t^{B^{'}}_{\;A^{'}}\varsigma
$.
 Since the element $\varsigma$ measures the degree
of elements on $M_{q}$, it is often called {\it dilaton element}
\cite{Schlieker/5/91,DAMTP/92-65}.

Thus q-Lorentz group and q-Minkowski space are given as   non-commutative
deformations of polynomial function algebras. This means that if we are
interested
in wave equations on quantum Minkowski space, we also have to dualise and
q-deform the notion of fields.
Classically, a  Lorentz field $\varphi$ on Minkowski space is a
left Lorentz group module morphism
$$
\begin{array}{rcl}
\varphi : M&\rightarrow &V\\
x&\mapsto & \varphi(x),
\end{array}
$$
where $V$ is some finite dimensional vector space and left Lorentz module.
If we denote the linear coordinate
functionals on $V$ by $\xi_{i}$, then $\varphi$ induces
a right ${\cal P}(L)$-comodule morphism
$$
\begin{array}{rcl}
\Phi: {\cal P}(V) &\rightarrow &{\cal P}(M)\\
\xi_{i} &\mapsto & \Phi(\xi_{i}) =  \xi_{i}\circ\varphi,
\end{array}
$$
which is also an algebra map. However, this algebra homomorphism property is
somewhat
accidental since  all algebras
are commutative, and we do not have any reason to expect this to hold in the
non-commutative case. Thus if $V_{q}=\bigoplus_{i=0}^{\infty} V_{i}$ is a
non-commutative deformation of ${\cal P}(V)$ as a
graded $\ast$-algebra, then we call a linear $\ast$-map
$$ \Phi: V_{1}\rightarrow M_{q}$$
from the linear component of $V_{q}$ to $M_{q}$
a {\it q-Lorentz field} if it is a right $\widetilde{\cal L}_{q}$-comodule
morphism,
irrespectively whether it extends as an algebra map  or not.

A solution to a Lorentz covariant wave equation ${\cal O}\varphi=0$, where
${\cal O}$ is some linear differential operator then induces  on the dual level
a solution to the  ${\cal P}(L)$-covariant equation
${\cal O}\Phi =0$. We take this as the general form of a covariant wave
equation also in
the q-deformed case.  But the necessity to extend the q-Lorentz group by the
dilaton
element suggests to  consider only
massless wave equations, since a mass term would destroy the scaling
invariance.
However, this is not too serious a restriction if we are interested in a toy
model
for Planck scale corrections to the geometry.

An outline of the paper is as follows. In section 2 we analyse differential
forms on
quantum Minkowski space and  introduce
the q-exterior derivative and the q-Hodge star operator. This leads to a
natural
definition of a q-coderivative, a q-Laplace-Beltrami operator and a q-Lie
derivative.
The crucial ingredient in these constructions is an $\widetilde{\cal
L}_{q}$-covariant
antisymmetrisation operation, which is developed in  2.1.
Section 3 then applies these results to the q-d'Alembert equation as the
simplest example
of a wave equation on quantum Minkowski space. Solutions of this equation
define
a conserved current, but unlike in the classical case this
current does not vanish for real solutions. We also give a family of q-deformed
plane
wave solutions to this equation. The fact that these plane waves exist only on
the q-light cone
gives further support to our claim that wave equations on q-Minkowski space
should be
massless. In section 4 we analyse the
q-Maxwell equations, and again  give a family of q-deformed
plane wave solutions.  We also discuss the q-gauge freedom of the q-Maxwell
equation.
Finally we then give an $SL_{q}(2,\cn)$-spinor
decomposition of the self-dual and anti-self-dual parts of the q-field strength
tensor.
 The spinor formulation of the q-Maxwell equation in the
last section also allows for a generalisation to arbitrary spin.

\section{Differential forms on quantum Minkowski space}

Since quantum Minkowski space has a braided coaddition  \cite {DAMTP/93-45}, we
can immediately apply the results of \cite{DAMTP/93-3}, where a braided
differential
calculus was developed. The key idea was to obtain braided differential
operators
$\partial^{a}$ by  `differentiating' the braided coaddition. The resulting
algebra of braided differential  operators ${\cal D} $ acts on quantum
Minkowski space
with an action $\alpha : {\cal D} \underline{\otimes} M_{q} \rightarrow M_{q}$
such that the  {\it braided Leibnitz rule}  holds \cite{DAMTP/93-3}
\begin{equation}\label{leibnitz}
\partial^{a} fg= (\partial^{a}f)g+\cdot\circ\Psi^{-1}_{L}(\partial^{a}\otimes
f)g.
\end{equation}
In \cite{DAMTP/93-3}  it was also shown that the braided differential operators
$\partial^{a}$ obey  the relations of $V({\bf R}_{M})$
\begin{equation}\label{brarel}
\partial^{a}\partial^{b}={\bf R}^{ \;\;ab}_{ M\; cd}\partial^{d}\partial^{c}.
\end{equation}
This means that we can define a covariant coaction
$\beta_{\cal D} : {\cal D}\rightarrow  {\cal D}\otimes \widetilde{\cal L}_{q}$
by $\partial^{a}\mapsto \partial^{b}\otimes\lambda^{a}_{\;b}\varsigma^{-1}$
and thus obtain a  coaction
$$\beta_{{\cal D} \underline{\otimes} M_{q}} : {\cal D} \underline{\otimes}
M_{q}
\rightarrow \left( {\cal D} \underline{\otimes} M_{q} \right) \otimes
\widetilde{\cal L}_{q},$$
which is an algebra map because of the covariance properties of the braided
tensor
product. This covariant coaction then
 makes the action $\alpha$ into an $\widetilde{\cal L}_{q}$-comodule morphism.
Note that the $\partial^{a}$  transform
with a scaling factor $\varsigma^{-1}$ inverse to the one for the coordinates,
as
appropriate for a derivative.

Thus there is well-developed theory of $\widetilde{\cal L}_{q}$-covariant
`q-partial derivatives' on quantum Minkowski
space, but q-exterior derivative, q-Hodge star operator and the q-coderivative
are far less well understood. In this section, we introduce these operations.

\subsection{Antisymmetrisers and the q-exterior algebra}\label{asy}

In the classical examples we would like to q-deform, fields are often forms
on Minkowski space. Thus we need a dual and q-deformed  generalisation of
the exterior algebra. We shall construct this algebra explicitly, and not
merely
define it abstractly as in \cite{Wess/92}. The essential
ingredient  in our approach  is an $\widetilde{\cal L}_{q}$-covariant
q-antisymmetrisation operation.
For this, we have to use a q-deformed notion of antisymmetry, where
we call an  $\widetilde{\cal L}_{q}$-tensor $T_{\ldots ab \ldots}$ {\it
q-antisymmetric}
 in
the adjacent indices $a$ and $b$ if
\begin{equation}\label{anti}
T_{\ldots ab \ldots}= - T_{\ldots cd \ldots}{\bf R}^{\;dc}_{L\;ab}.
\end{equation}
Here we use an index notation for tensors, where $T_{ab}$ and $T^{ab}$, etc.
denote  any
elements
of right  $\widetilde{\cal L}_{q}$-comodules, which transform  as
$T_{ab}\mapsto  T_{cd}\otimes \lambda^{c}_{\;a}\lambda^{d}_{\;b}\varsigma^{n}$
and
$T^{ab} \mapsto T^{cd}\otimes
S\lambda^{a}_{\;c}S\lambda^{b}_{\;d}\varsigma^{m}$,
where $S$ denotes the antipode in ${\cal L}_{q}$.
We do not require a tensor  to have a specific $\varsigma$-scaling
property, and therefore $n$ and $m$ can be any integers.

As shown in  \cite{DAMTP/93-45}, we can use the q-metric $g^{ab}$ and its
inverse
to raise and lower indices. Due to the relation between the R-matrix and the
q-metric
\cite{DAMTP/93-45}
\begin{equation}\label{raise}
{\bf R}^{\;\;kl}_{L \;ef}=g_{pf}g_{qe}
{\bf R}^{\;\;qp}_{L\;ab}g^{ak}g^{bl},
\end{equation}
this raising and lowering of indices preserves the  q-antisymmetry of tensors.
If for example
a tensor $T_{\ldots ab \ldots}$ is q-antisymmetric in $a$ and $b$ then the
tensor with upper indices
$T^{\ldots ab \ldots}=T_{\ldots ij \ldots} \ldots g^{ia}g^{jb}\ldots$ obeys
$$
T^{\ldots ab \ldots}= -{\bf R}^{\;ab}_{L\;dc} T^{\ldots cd \ldots},
$$
A tensor is called $q-symmetric$ in
$a$ and $b$ if
$$
T_{\ldots ab\ldots}=   T_{\ldots cd \ldots}{\bf R}^{\;dc}_{M\;ab}.
$$
Again, this translates into the corresponding formula for upper indices by
virtue
of a relation of the type (\ref{raise}) for the matrix ${\bf R}_{M}$. Note that
 we are using two
different R-matrices for the definition of q-symmetry and q-antisymmetry.

Since   ${\bf R}_{L}$ and ${\bf R}_{M}$ obey the relation (\ref{quasiheck}),
one might suspect that $(P{\bf R}_{M}-1)$ would be a good candidate for a
q-antisymmetriser.
However, this operator is not a projector, and it is also not quite clear how
to obtain higher
antisymmetrisers. We shall therefore take a different approach.
Recall that in the classical case, the space of totally antisymmetric tensors
of valence
four is one-dimensional, and one can choose  a basis vector
$\varepsilon_{abcd}$ with $\varepsilon_{1234}=1$, which defines a projector
(antisymmetriser)
$\frac{1}{4!}\varepsilon^{dcba}\varepsilon_{efgh}$ onto this one-dimensional
space.
 By successively contracting indices, one obtains the lower antisymmetrisers.
The q-deformed case is quite similar:

\begin{lemma}
Up to a factor, there is exactly one complex valued tensor
$\varepsilon_{abcd}$,
which is  totally q-antisymmetric in any two adjacent indices.
\end{lemma}

{\bf Proof.} One can show explicitly that the system of linear equations
$\varepsilon_{abcd}=-\varepsilon_{ijcd}{\bf R}^{\;ji}_{L\;ab}
=-\varepsilon_{aijd}{\bf R}^{\;ji}_{L\;bc}
=-\varepsilon_{abij}{\bf R}^{\;ji}_{L\;cd}$
has a one-dimensional solution space.
The non-zero entries of $\varepsilon_{abcd}$ in the normalisation
$\varepsilon_{1234}=1$ are:
$$\begin{array}{rclcrclcrclcrcl}
\varepsilon_{1234}&=&1&&\varepsilon_{1243}&=&-q^{-2}&&\varepsilon_{1324}&=&-1&&
\varepsilon_{1342}&=& q^{2}\\
\varepsilon_{1414}&=& 1-q^{2}&& \varepsilon_{1423}&=&1&&
\varepsilon_{1432}&=& -1&& \varepsilon_{1444}&=& 1-q^{-2}\\
\varepsilon_{2134}&=& -1&&
\varepsilon_{2143}&=& q^{-2}&& \varepsilon_{2314}&=& 1&& \varepsilon_{2341}&=&
-1 \\
\varepsilon_{2413}&=& -q^{-2}&&
\varepsilon_{2431}&=&q^{-2}&&\varepsilon_{2434}&=& q^{-2}-1&&
\varepsilon_{3124}&=& 1\\
 \varepsilon_{3142}&=& -q^{2}&&\varepsilon_{3214}&=& -1&&
\varepsilon_{3241}&=& 1 && \varepsilon_{3412}&=& q^{2}\\
 \varepsilon_{3421}&=& -q^{2} &&
\varepsilon_{3424}&=& 1-q^{2}&& \varepsilon_{4123}&=& -1&&\varepsilon_{4132}&=&
1\\
\varepsilon_{4141}&=&q^{2}-1&& \varepsilon_{4144}&=& q^{-2}-1&&
\varepsilon_{4213}&=& 1&&
\varepsilon_{4231}&=& -1\\
\varepsilon_{4243}&=& 1-q^{-2}&&\varepsilon_{4312}&=& -1&&
\varepsilon_{4321}&=& 1&& \varepsilon_{4342}&=& q^{2}-1\\
 \varepsilon_{4414}&=& 1-q^{-2} &&\varepsilon_{4441}&=& q^{-2}-1
\end{array}$$
\hfill $\Box$

Unlike the $\epsilon$'s for $SU_{q}(n)$ where only the $\pm 1$ entries of
$\epsilon$ are changed to powers of $q$ and all zero entries remain unchanged,
we here obtain  non-zero entries like $\varepsilon_{4441}$, etc.
In terms of  this q-antisymmetric tensor, we define   q-antisymmetrisers
by successively contracting indices:

\begin{defi}\label{quanti}
Let $T_{\ldots a_{1}\ldots a_{n} \ldots}$ be an  $\widetilde{\cal
L}_{q}$-tensor.
We define its {\bf q-antisymmetrisation} in the adjacent indices $ a_{1}\ldots
a_{n} $ by
$$T_{\ldots [a_{1}\ldots a_{n}] \ldots}:=T_{\ldots c_{1}\ldots c_{n} \ldots}\,
 {\cal A}^{\;\;c_{1}\ldots c_{n}}_{ \{n\} \;a_{1}\ldots a_{n}}$$
for $n<5$ and zero otherwise. The {\bf q-antisymmetrisers} $ {\cal A}_{\{k\} }
$ are defined as
\begin{equation}
\begin{array}{rclcrcl}
 {\cal A}^{\;\;abcd}_{\{4\}\;efgh}&:=& \frac{1}{ n_{4}}
\,\varepsilon^{dcba}\varepsilon_{efgh}&&
 {\cal A}^{\;\;abc}_{\{3\}\;fgh}&:=& -\frac{1}{ n_{3}}
\,\varepsilon^{dcba}\varepsilon_{dfgh}\\
\\
 {\cal A}^{\;\;ab}_{\{2\}\;gh}&:=& \frac{1}{ n_{2}}
\,\varepsilon^{dcba}\varepsilon_{cdgh}&&
 {\cal A}^{\;\;a}_{\{1\}\;h}&:=& -\frac{1}{ n_{1}}
\,\varepsilon^{dcba}\varepsilon_{bcdh},
\end{array}
\end{equation}
where the normalisation factors $n_{k}$
$$
\begin{array}{rclcrcl}
n_{1}&=& 2(1+q^{2}+q^{4}), && n_{2}&=& (1+q^{2})(1+q^{2}) \\
n_{3}&=&n_{1},&& n_{4}&=&q^{-2} 2 (1+q^{2}+q^{4})(1+q^{2})(1+q^{2})
\end{array}
$$
are a q-deformation of $ (4-k)!k! $.
\end{defi}

The q-antisymmetrisation of a tensor is clearly q-antisymmetric in the sense
of (\ref{anti}), but also has the other properties one might expect:

\begin{prop}
By explicit calculation, one can show:
\begin{enumerate}
\item The antisymmetrisers ${\cal A}_{\{k\} } $ are projectors:
\begin{equation}\label{proj}
{\cal A}^{2}_{\{k\} } = {\cal A}_{\{k\} } ,\;\;\;\; \mbox{i.e.}\;\; \;\;
T_{\ldots [[a_{1}\ldots a_{n}]] \ldots}=T_{\ldots [a_{1}\ldots a_{n}] \ldots}.
\end{equation}
for $k=1\ldots 4$.
\item Lower dimensional q-antisymmetrisers cancel on higher dimensional ones:
\begin{equation}\label{cancel}
T_{\ldots [a_{1}\ldots [a_{k}\ldots a_{l}]\ldots a_{n}] \ldots}
=T_{\ldots [a_{1}\ldots a_{n}] \ldots}.
\end{equation}
\item The one-dimensional projector is trivial:
 $${\cal A}_{\{1\} }  = 1,\;\;\;\; \;\;\;\mbox{i.e.}\;\; \;\;\;T_{\ldots [a]
\ldots}=T_{\ldots a \ldots}$$
\item
There are two invertible matrices $B$ and $B^{'}$ such that
\begin{equation} \label{nohe}
{\cal A}_{\{2\} } = (P{\bf R}_{M} - 1)B=B^{'} (P{\bf R}_{M} - 1).
\end{equation}
I.e., the two-dimensional antisymmetriser ${\cal A}_{\{2\} }$ factors through
$(P{\bf R}_{M}-1)$.
\end{enumerate}
\end{prop}

In addition to (\ref{proj}), one usually requires a linear operator to be
Hermitean
with respect to a given inner product before calling it `projector'. We can
show
something similar in our case, e.g. ${\cal A}_{\{2\}}$ is `Hermitean' with
respect
to the q-deformed metric in the sense that
$${\cal A}^{\;\; ab}_{\{2\} cd} = g_{cj}  g_{di} {\cal A}^{\;\; ij}_{\{2\} kl}
g^{al}g^{bk}$$
and similar relations for the other q-antisymmetrisers, but these properties
are not
important for our purposes and we therefore do not discuss them in detail.
One might regard (\ref{proj}) as the minimal requirement for a
q-antisymmetrisation
operation to make sense, but  for the following  all the other properties are
needed
as well, in particular (\ref{nohe}). This relation ensures that q-symmetric
tensors are in the kernel of
the q-antisymmetrisers. To make this point explicit, note that (\ref{cancel})
and
(\ref{nohe}) imply:

\begin{cor}\label{ococ}
If an $\widetilde{\cal L}_{q}$-tensor $T_{\ldots a_{1}\ldots a_{k}\ldots }$ is
q-symmetric
in two adjacent indices $a_{i}$ and $a_{i+1}$ then
 \begin{equation}\label{coco}
T_{\ldots [a_{1}\ldots a_{i}a_{i+1}\ldots a_{k}]\ldots } =0.
\end{equation}
\end{cor}

Thus although q-symmetry and q-antisymmetry are defined in terms of two
different
R-matrices, the two notions are compatible in this sense.
The relation (\ref{coco}) shall be of
importance in section \ref{extal}, where we introduce
the external derivative $d$ and show that $d^{2}=0$. Note also that one can
easily
define a q-symmetriser on two adjacent indices
as $S_{\{2\}} = \frac{1}{2}(1-A_{\{2\}})$,
but in our  non-Hecke case it not quite clear how to obtain higher
q-symmetrisers.

So far, we have used the index notation for q-antisymmetrised tensors  without
addressing the problem of $\widetilde{\cal L}_{q}$-covariance of the
q-antisymmetrisation. This question is answered by the following proposition:

\begin{prop} \label{comm}
The coaction $\beta$ by the q-Lorentz group commutes with the operation of
q-antisymmetrisation. Symbolically,
$$\beta\circ [\;\;\;]=[\;\;\;]\circ\beta .$$
This means in particular that the property
of q-antisymmetry is covariant under the coaction by $\widetilde{\cal L}_{q}$.
\end{prop}

{\bf Proof.}
We need to show that monomials  of generators of  ${\cal L}_{q}$
commute with the q-antisymmetriser ${\cal A}_{\{n\}}$.
The generators of the q-Lorentz group  obey \cite{DAMTP/93-45}
$$\lambda^{a}_{\;e}\lambda^{b}_{\;f}\lambda^{c}_{\;g}\lambda^{d}_{\;h}
\varepsilon^{hgfe}=n_{4}^{-1} \, \varepsilon^{dcba}
\varepsilon_{mnop}\lambda^{m}_{\;e}\lambda^{n}_{\;f}\lambda^{o}_{\;g}
\lambda^{p}_{\;h}
\varepsilon^{hgfe}.$$
This implies for the q-antisymmetriser ${\cal A}_{\{ 2\} }$:
$$
\begin{array}{rcl}
\lambda^{a}_{\;e}\lambda^{b}_{\;f}
\varepsilon^{dcfe}\varepsilon_{cdkl}
&=&\lambda^{a}_{\;e}\lambda^{b}_{\;f} \delta^{c}_{\;g}\delta^{d}_{\;h}
\varepsilon^{hgfe}\varepsilon_{cdkl}\\
&=&\lambda^{a}_{\;e}\lambda^{b}_{\;f}\lambda^{n}_{\;g}  \lambda^{m}_{\;h}
\varepsilon^{hgfe}
S^{-1}\lambda^{d}_{\;m}S^{-1}\lambda^{c}_{\;n}\varepsilon_{cdkl}\\
&=&n_{4}^{-1} \,\varepsilon^{mnba}\varepsilon_{opqr}
\lambda^{o}_{\;e}\lambda^{p}_{\;f}\lambda^{q}_{\;g}  \lambda^{r}_{\;h}
\varepsilon^{hgfe}
S^{-1}\lambda^{d}_{\;m}S^{-1}\lambda^{c}_{\;n}\varepsilon_{cdkl}\\
&=&n_{4}^{-1} \,\varepsilon^{mnba}
S^{-1}\lambda^{d}_{\;m}S^{-1}\lambda^{c}_{\;n}
\varepsilon_{opqr}
\lambda^{o}_{\;e}\lambda^{p}_{\;f}\lambda^{q}_{\;g}  \lambda^{r}_{\;h}
\varepsilon^{hgfe}\varepsilon_{cdkl}\\
&=&\varepsilon^{mnba}
S^{-1}\lambda^{d}_{\;m}S^{-1}\lambda^{c}_{\;n}
\varepsilon_{opqr}
\lambda^{o}_{\;c}\lambda^{p}_{\;d}\lambda^{q}_{\;k}  \lambda^{r}_{\;l}\\
&=&\varepsilon^{mnba}\varepsilon_{nmqr}\lambda^{q}_{\;k}  \lambda^{r}_{\;l},
\end{array}
$$
and similar for ${\cal A}_{\{3\} }$ and ${\cal A}_{\{4\} }$. The case of ${\cal
A}_{\{1\} }$ is trivial.
\hfill $\Box$

In terms of  these q-antisymmetrisers,
one can now define a q-deformation of the wedge product of copies
of $M_{q}$, realised in the braided
tensor product. Explicitly, let $M_{q} \wedge \ldots\wedge M_{q}$
be the subalgebra of
$M_{q} \underline{\otimes}\ldots\underline{\otimes}M_{q}$
generated by
$$
\begin{array}{rcl}
dx_{a_{1}}\wedge \ldots\wedge dx_{a_{n}}
&:=&dx_{[a_{1}}\underline{\otimes}\ldots\underline{\otimes}dx_{a_{n}]}\\
&=&dx_{b_{1}}\underline{\otimes}\ldots\underline{\otimes}dx_{b_{n}}\,
 {\cal A}^{\;\;b_{1}\ldots b_{n}}_{\{ n\} \;a_{1}\ldots a_{n}},
\end{array}
$$
where $dx$ denotes a copy of the generators of $M_{q}$.
This means that the linear component
$$\Lambda_{n}:=(\underbrace{
M_{q} \wedge \ldots\wedge M_{q}}_{n}
)_{1}$$
is   spanned by the totally q-antisymmetric
$$
{\bf e}_{a_{1} \ldots a_{n}}:=
dx_{a_{1}}\wedge \ldots\wedge dx_{a_{n}}
$$
with relations of the form (\ref{anti}).
We note that
$\Lambda_{n}$ is very similar
to the general form of the  exterior algebras discussed in
\cite{WessZumino/90/1}
for  the special case
where all R-matrices are of Hecke type.
In the framework set out in section \ref{prelim},
we now  define a  {\it p-form on quantum Minkowski space} as an
$\widetilde{\cal L}_{q}$-comodule morphism
$$ {\bf w}: \Lambda_{p} \rightarrow M_{q}.$$
For $p=0$, we define
$\Lambda_{0}:=\cn$ as the linear component of ${\cal P}(\rn)$,
the  algebra of complex valued polynomial functions on $\rn$, and choose a
basis
 element  ${\bf e}$.

Over the ring of all $\widetilde{\cal L}_{q}$-scalars in $M_{q}$,
$p$-forms  form a linear space, which is denoted by $\Omega_{p}$.
A simple analysis of the rank of the antisymmetrisers ${\cal A}_{\{ p\} }$
shows that
these spaces
have the  dimensions $1,4,6,4,1$ for $p=0,1,2,3,4$, and dimension 0 for $p>4$.
Thus the dimensions are the same as in the classical case.
The one-dimensional space $\Omega_{4}$ is spanned by the
{\it top form} $\mbox{\boldmath $\varepsilon$}$ defined as
\begin{equation}\label{topform}
\mbox{\boldmath $\varepsilon$}: {\bf e}_{a_{1} \ldots a_{4}}
\mapsto  \varepsilon_{a_{1} \ldots a_{4}}.
\end{equation}

As linear maps, $p$-forms ${\bf w}$ are determined by their value on
the  elements ${\bf e}_{a_{1} \ldots a_{p}}$ of $\Lambda_{p}$, but we also
have:

\begin{prop}
All p-forms on quantum Minkowski space are of the form
$$
{\bf w}({\bf e}_{a_{1} \ldots a_{p}})
= w_{[ a_{1}\ldots a_{p}] },
$$
for some $w_{a_{1}\ldots a_{p}}\in M_{q}$.
\end{prop}

{\bf Proof.}
The proposition is tantamount to claiming that all totally q-antisymmetric
tensors
are in the image of the q-antisymmetrisers. Hence the proposition can be proved
by
verifying that the dimension of totally q-antisymmetric tensors over the ring
of
$\widetilde{\cal L}_{q}$-scalars in $M_{q}$ coincides with the ranks of the
q-antisymmetrisers ${\cal A}_{\{ k\} }$.\hfill $\Box$

As in the classical case, the space of forms on quantum Minkowski space
$$\Omega :=\bigoplus_{p=0}^{4} \Omega_{p},$$
can be equipped with an algebra structure.
Given a $p$-form $\bf w$ and an $r$-form $\bf v$,   their
{\it q-wedge product} ${\bf w}\wedge {\bf v} \in \Omega_{p+r}$ is defined as:
\begin{equation}\label{wdef}
{\bf w}\wedge {\bf v}:
{\bf e}_{a_{1} \ldots a_{p+r}}\mapsto w_{[a_{1}\ldots a_{p}}v_{a_{p+1}\ldots
a_{p+r}]}.
\end{equation}
By virtue of  (\ref{cancel}), this bilinear operation `$\wedge$' defines an
associative
algebra structure on
$\Omega $, the {\it q-exterior algebra}, with   identity  ${\bf 1}\in
\Omega_{0}$
given by ${\bf 1}:  {\bf e}  \mapsto  1$.

\subsection{The q-exterior derivative}\label{extal}

In this section we define an q-exterior derivative on forms on q-Minkowski
space.
Essential building blocks are  the braided differential operators
$\partial^{a}$
and  the  q-antisymmetrisers.

\begin{defi}
The {\bf q-exterior derivative} $d:\Omega_{p}\mapsto\Omega_{p+1}$  is defined
as
$$d{\bf w}({\bf e}_{a_{1} \ldots a_{p+1}})
:= \partial_{[a_{1}}w_{a_{2}\ldots a_{p+1}]}.$$
\end{defi}

Forms whose q-exterior derivative vanishes are called
{\it closed} and forms which are themselves q-exterior derivatives are said
to be {\it exact}. The crucial test for a definition of a `q-deformed
exterior derivative' is whether exact forms are closed. In
\cite{WessZumino/90/1},
for example, this could be shown only in the (trivial) Hecke case.
Due to the careful construction of the q-antisymmetrisers, we obtain in our
non-Hecke case:
\begin{prop}\label{exclo}
Exact forms on q-Minkowski space are closed:
$$d^{2}=0.$$
\end{prop}

{\bf Proof.}
Relation (\ref{cancel}) implies for exact forms $d{\bf w}$:
$$
\begin{array}{rcl}
 d^{2} {\bf w}({\bf e}_{a_{1} \ldots a_{p+2}})
&= &\partial_{[a_{1}}\partial_{a_{2}}w_{a_{3}\ldots a_{p+1}]}\\
&= &\partial_{[[a_{1}}\partial_{a_{2}]}w_{a_{3}\ldots a_{p+1}]}\\
&=&0
\end{array}
$$
Here we used that  braided differential operators $\partial_{a}$  obey the
relations
of $V^{'}({\bf R}_{M})$ \cite{DAMTP/93-3},  i.e.
$\partial_{a_{1}}\partial_{a_{2}}$ is q-symmetric  and thus
$\partial_{[a_{1}}\partial_{a_{2}]}=0$ by virtue of corollary \ref{ococ}.
\hfill  $\Box$

Using the recent results on q-integration by A. Kempf and S. Majid
\cite{DAMTP/94-7}
one might also be able to prove a q-Poincar\'{e} lemma for q-Minkowski space,
which
would then imply that all closed forms are exact.

 As a consequence of the braided Leibnitz rule (\ref{leibnitz}) we find for the
action
of the q-exterior derivative $d$ on wedge products of forms:

\begin{cor}\label{leiwed}
The q-exterior derivative $d$ acts as
$$d{\bf w}\wedge  {\bf v} =
(d{\bf w})\wedge  {\bf v}+(-1)^{p} {\bf w}\wedge d{\bf v} $$
on wedge products ${\bf w}\wedge  {\bf v}$, where $\bf w$ is a $p$-form.
\end{cor}

{\bf Proof.}
The crucial point is that the inverse braiding brings up R-matrices, which
cancel
on the q-antisymmetriser because of the symmetry property (\ref{anti}).
We prove the corollary only for 1-forms  ${\bf w}$, the general case follows
immediately by using the hexagon identity for the braiding $\Psi$.
Thus let ${\bf w}$ be a 1-form and $\bf v$ a p-form. On  $\Lambda_{p+2}$
we  have by virtue of  (\ref{leibnitz}) and
the q-antisymmetry of $\varepsilon^{abcd}$:
$$
\begin{array}{rcl}
\partial_{[a_{1}}w_{a_{2}}v_{a_{3}\ldots
a_{p+2}]}
&=&(\partial_{[a_{1}}w_{a_{2}}) v_{a_{3}\ldots
a_{p+2}]}+ \cdot\circ\Psi^{-1}(\partial_{[a_{1} }
\otimes w_{a_{2}}) v_{a_{3}\ldots a_{p+2}]}\\
&=& (\partial_{[a_{1}}w_{a_{2}}) v_{a_{3}\ldots
a_{p+2}]}+ w_{c}\otimes \partial_{d} {\bf R}^{-1 cd}_{L\; [a_{2}a_{1}}
v_{a_{3}\ldots a_{p+2}]}\\
&=&  (\partial_{[a_{1}}w_{a_{2}}) v_{a_{3}\ldots
a_{p+2}]}+ w_{[a_{2}}\otimes \partial_{a_{1}}  v_{a_{3}\ldots a_{p+2}]}
\end{array}
$$
Here we used the inverse braiding  $\Psi^{-1}(\partial_{ a_{1} }
\otimes w_{a_{2}}) = w_{c}\otimes \partial_{d} {\bf R}^{-1 cd}_{L\;
a_{2}a_{1}}$
(See \cite[Prop. 3.2]{DAMTP/92-12} for a useful list of braidings between
various standard
algebras).
\hfill $\Box$

\subsection{q-Hodge star operator}

The other operation on  forms  on q-Minkowski space
one can define with the tools at hand is the
q-Hodge  star operator. It is  defined
in terms of the metric $g^{ab}$ and the tensor $\varepsilon_{abcd}$.
\begin{defi}
The {\bf q-Hodge star operator}
$\ast: \Omega_{p} \rightarrow \Omega_{4-p}$
is defined by
$$
\begin{array}{rcl}
^{\ast}{\bf w} ({\bf e}_{a_{1} \ldots a_{4-p}})
&=& n_{p}^{-1/2}\varepsilon_{a_{1}\ldots a_{4-p}b_{1}\ldots b_{p}}
g^{b_{1} c_{p}} \ldots
g^{b_{p} c_{1}}w_{[c_{1}\ldots c_{p}]}\\
\\
&=:&w_{[c_{1}\ldots c_{p}]}H^{c_{1}\ldots c_{p}}_{\{ p\} \;a_{1}\ldots
a_{4-p}},
\end{array}
$$
with the normalisation factor $n_{0}:=n_{4}$.
\end{defi}

Again, this definition will only be justified if we can show some non-trivial
properties.
We shall use the q-Hodge star to define a q-coderivative $\delta $ and a
q-Laplace-Beltrami
operator $\mbox{\boldmath $\Delta$}$. For these operators to have reasonable
properties, it is necessary for the square of the q-Hodge star  to have a
sufficiently well-behaved form. It turns out that  we recover exactly the
classical result:

\begin{prop}\label{hodge2}
The square of the q-Hodge star operator on $p$-forms
is given by:
\begin{equation}\label{hosqu}
\ast\circ\ast = (-1)^{p(4-p)}.
\end{equation}
Furthermore, one finds
 $$^{\ast}\mbox{\boldmath $\varepsilon$}={\bf 1},\;\;\; ^{\ast}{\bf 1}=
\mbox{\boldmath $\varepsilon$}$$
for the special cases of the top form
$\mbox{\boldmath $\varepsilon$}$ and the identity form  $\bf 1$.
\end{prop}

{\bf Proof.}
By explicit calculation, one can verify the following relations between the
q-antisymmetrisers ${\cal A}_{\{k\} }$ and the matrices $H_{\{k\} }$, which
implement the q-Hodge star operation:
\begin{equation}\label{ho1}
\begin{array}{rcl}
H_{\{0\}}H_{\{4\}}&=&1\\
-H_{\{1\}}H_{\{3\}}&=&{\cal A}_{\{1\}}\\
H_{\{2\}}H_{\{2\}} &=&{\cal A}_{\{2\}}\\
-H_{\{3\}}H_{\{1\}}&=&{\cal A}_{\{3\}}\\
H_{\{4\}}H_{\{ 0\}}&=&{\cal A}_{\{4\}}
\end{array}
\end{equation}
Together with (\ref{proj}), these relations imply the proposition. \hfill $\Box
$

Since we are working in a `spinorial basis' we do not obtain an additional
$(-1)$-factor on the right hand side of relation (\ref{hosqu}), as in the case
of
an `$x,y,z,t$-basis'. Another property of the q-Hodge star operator   is that
one can
`shift' the q-Hodge star operator in the q-wedge product of two $p$-forms:

\begin{lemma}\label{shiftho}
If $\bf w$ and $\bf v$ are both $p$-forms, then
$$^{\ast}{\bf w}\wedge {\bf v}= (-1)^{p} {\bf w}\wedge ^{\ast}{\bf v}.$$
\end{lemma}

{\bf Proof.}
It is sufficient to verify the relations
$$
\begin{array}{rcl}
-H^{\;a}_{\{1\}bcd}\,{\cal A}^{\;bcde}_{\{4\} klmn}
&=&  H^{\;e}_{\{1\}bcd}\,{\cal A}^{\;abcd }_{\{4\} klmn}\\
H^{\;ab}_{\{2\} cd}\;\;{\cal A}^{\;cdef}_{\{4\} klmn}
&=& H^{\;ef}_{\{2\} cd}\;\;{\cal A}^{\;abcd}_{\{4\} klmn}\\
-H^{\;abc}_{\{3\}  d}\;\;\;{\cal A}^{\;defg}_{\{4\} klmn}
&=& H^{\;efg}_{\{3\}  d}\;\;\;{\cal A}^{\;abcd}_{\{4\} klmn}
\end{array}
$$
which establishes the lemma. \hfill $\Box$

\subsection{q-Coderivative, q-Laplace-Beltrami operator  and the q-Lie
derivative}

Now that we are given both a well-behaved exterior derivative
and a q-Hodge star operator, it is straightforward to define a suitable
notion of q-coderivative, q-Laplace-Beltrami operator and q-Lie derivative.

\begin{defi}
The {\bf q-coderivative} $\delta : \Omega_{p}\rightarrow\Omega_{p-1}$
and the {\bf q-Laplace-Beltrami operator}
$\mbox{\boldmath $\Delta$}: \Omega_{p} \rightarrow \Omega_{p}$
on $p$-forms on quantum Minkowski space are defined as
$$\delta := \,^{\ast}  d \, ^{\ast} ,\;\;\;\;\;\; \mbox{\boldmath
$\Delta$}:=\delta  d
 + d \delta.$$
\end{defi}

Forms $\bf w$ on quantum Minkowski space which satisfy $\delta {\bf w}=0$ are
called {\it co-closed}, and forms which are themselves q-coderivatives are
called
{\it co-exact}. As a corollary of proposition  \ref{hodge2} and  proposition
\ref{exclo}, one finds:

\begin{cor}
Co-exact forms on quantum Minkowski space are  co-closed:
$$\delta ^{2}=0.$$
\end{cor}

Thus, although $d$, $\delta$ and $\mbox{\boldmath $\Delta$}$ are defined in
terms of
deformed antisymmetrisers and
differential operators on a non-commutative space, their abstract properties
resemble very much the classical case. It is straightforward to verify:
\begin{equation}
\begin{array}{rclcrcl }\label{crrels}
d \mbox{\boldmath $\Delta$}&=& \mbox{\boldmath $\Delta$} d
&&
\delta \mbox{\boldmath $\Delta$}&=&\mbox{\boldmath $\Delta$}\delta \\
\mbox{\boldmath $\Delta$}^{\ast} &=& ^{\ast}  \mbox{\boldmath $\Delta$}&&
\delta ^{\ast} &=& (-1)^{p}\,^{\ast}  d \\
 ^{\ast} \delta &=& (-1)^{p+1}  d  ^{\ast} &&
d \delta  ^{\ast} &=& ^{\ast}\delta d  \\
^{\ast} d \delta &=& \delta d ^{\ast} &&
\mbox{\boldmath $\Delta$}^{\ast} &=& ^{\ast}  \mbox{\boldmath $\Delta$}\\
\end{array}
\end{equation}
Further, and less trivial, properties will be given in the following
sections. In particular we will analyse the explicit action of these operators
on zero
and 1-forms, which is  of  interest to physical applications.

The q-Hodge star operator also enables us to generalise the idea of a Lie
derivative.
For this purpose, we introduce a  {\it q-inner product} on the q-exterior
algebra
$\Omega$ as a bilinear map
$(\;\;,\,\,): \Omega_{p} \times \Omega_{r}  \rightarrow   \Omega_{p-r}$
defined by
$$  ({\bf w},{\bf v}) := i_{\bf v} {\bf w} :=
\,^{\ast}(  {\bf v} \wedge ^{\ast^{-1}} ({\bf w} )),$$
where ${\bf w}$ is a $p$-form.
The q-inner product is `transposed' to the q-wedge product in the sense that
$$({\bf v}\wedge {\bf w}, {\bf u})= i_{{\bf v}\wedge {\bf w}}{\bf u}=
i_{{\bf v}}( i_{{\bf w}} {\bf u} )=  ({\bf w}, i_{\bf v}{\bf u}),$$
and one can also show the formulae
$$
\begin{array}{rcl}
^{\ast}{\bf w} &=& i_{\bf w} \mbox{\boldmath $\varepsilon$}\\
\\
\delta i_{\bf v} {\bf w} &=& i_{\bf v}\delta {\bf w} + (-1)^{p} i_{d{\bf
v}}{\bf w},
\end{array}
$$
where ${\bf v}$ is a $p$-form. Furthermore, lemma \ref{shiftho} implies
$$({\bf v}, {\bf w})=i_{\bf v}{\bf w} = i_{^{\ast}{\bf v}} \, ^{\ast}{\bf w}=
(^{\ast}{\bf v},
^{\ast}{\bf w}),$$
for any two $p$-forms $\bf v$ and $\bf w$.
In terms of this q-inner product, we now introduce:

\begin{defi}
Let $\bf v$ be a 1-form. The {\bf q-Lie derivative }
with respect to $\bf v$ is defined as:
$$L_{\bf v} := i_{\bf v}\circ d + d \circ  i_{\bf v},$$
and is obviously a linear map $L_{\bf v}: \Omega_{p}\rightarrow \Omega_{p}$.
\end{defi}

The q-Lie derivative commutes with the q-exterior derivative
$$L_{\bf v} d = dL_{\bf v} ,$$
and we also have
$$L_{{\bf f}\wedge {\bf v}} {\bf w}={\bf f}\wedge L_{\bf v}{\bf w} +d{\bf
f}\wedge i_{\bf v}{\bf w}$$
for zero forms $\bf f$ and 1-forms $\bf v$.
For the action of the q-Lie derivative on zero forms on q-Minkowski space, we
find:
\begin{prop}
Let ${\bf f}\in \Omega_{0}$  and let ${\bf v}\in\Omega_{1}$. Then
the action of $L_{\bf v}$ on ${\bf f}$ is given by
$$L_{\bf v} {\bf f} : {\bf e} \mapsto v^{a} \partial_{a} f.$$
\end{prop}

{\bf Proof.}
First note that $L_{\bf v} {\bf f} = i_{\bf v} d {\bf f}$, since ${\bf f}$ is a
zero form. Then
 show by explicit calculation, that
\begin{equation}\label{xxx}
H^{a}_{\{1\} cde}H_{\{4\}}^{bcde} = - g^{ba}.
\end{equation}
Therefore
$$
\begin{array}{rcl}
L_{\bf v} {\bf f} ({\bf e})
&=&
 -v_{b} \partial_{a} f H^{a}_{\{1\} cde}{\cal A}^{bcde}_{\{4\}
klmn}H_{\{4\}}^{klmn}\\
&=&-v_{b} \partial_{a} f H^{a}_{\{1\} cde}H_{\{4\}}^{bcde}\\
&=&v_{b}\partial_{a} g^{ba} f,
\end{array}
$$
 \hfill $\Box$

\section{q-Scalar field}

\subsection{The q-d'Alembert equation}

The simplest case  of a wave equation on q-Minkowski space is the
q-d'Alembert equation,
where fields are 0-forms $\mbox{\boldmath $\varphi$}$ and the wave equation
is given by the q-Laplace-Beltrami operator:

\begin{defi}
A solution of the q-d'Alembert equation is a 0-form $\mbox{\boldmath
$\varphi$}$ such that
\begin{equation}\label{dalembert}
 \mbox{\boldmath $\Delta$}\mbox{\boldmath $\varphi$} = 0.
\end{equation}
\end{defi}

This equation can   be written less abstractly in terms of the braided
differential operators and the value $\varphi$ on $\bf e$ of
$\mbox{\boldmath $\varphi$}$.

\begin{prop}\label{obvi}
Equation (\ref{dalembert}) is equivalent to
\begin{equation}\label{dal2}
\Box\varphi =0,
\end{equation}
where $\Box := \partial_{a}\partial_{b}g^{ab}$ is the
{\bf q-d'Alembert operator}.
\end{prop}

{\bf Proof.}
Since $^{\ast}  \mbox{\boldmath $\varphi$}$ is a 4-form,
$d\delta  \mbox{\boldmath $\varphi$}$ vanishes and thus
$ \mbox{\boldmath $\Delta$}  \mbox{\boldmath $\varphi$} = \delta d
\mbox{\boldmath $\varphi$}$.
With relation (\ref{xxx}), we find:
$$
\begin{array}{rcl}
0 &= & \delta d \mbox{\boldmath $\varphi$} ({\bf e})\\
&=& \partial_{f}\partial_{a} \varphi H^{\;a}_{\{1\} cde}
{\cal A}^{\;fcde}_{\{4\} klmn} H^{klmn}_{\{4\}}\\
&=& \partial_{f}\partial_{a} \varphi H^{\;a}_{\{1\} cde} H^{fcde}_{\{4\}}\\
&=& \partial_{f}\partial_{a}g^{fa} \varphi
\end{array}
$$
which proves the equivalence of (\ref{dalembert}) and (\ref{dal2}).\hfill
$\Box$

In the form (\ref{dal2}), it is easy to see that the q-d'Alembert equation is
 $\widetilde{\cal L}_{q}$-covariant.
Keeping in mind  the various transformation properties, one can show that
the action $\alpha$ of the operator $\Box$
commutes with the coaction by $\widetilde{\cal L}_{q}$:
$$
\beta_{M_{q}}\circ \alpha  \circ (\Box\otimes \varphi ) =
\alpha \circ \beta_{{\cal D}\otimes M_{q}}\circ (\Box\otimes  \varphi ).
$$
One could also write down a {\it q-Klein Gordon equation} of the form
$$ (\Box + m^{2})\varphi =0,$$
but this equation would only be  $\widetilde{\cal L}_{q}$-covariant if the
`mass' $m$
transformed as $m\mapsto m\otimes\varsigma^{-1}$, i.e. not as an
$\widetilde{\cal L}_{q}$-scalar. One might argue that this transformation
property
in itself is not necessarily  harmful, but the results of the next section on
plane
wave solutions seem to suggest to us that   $\widetilde{\cal L}_{q}$-covariant
wave equations on $M_{q}$ are inherently massless.

The requirement for $\varphi$ to be a 0-form corresponds classically to a
restriction to real-valued solutions. A {\it complex valued solution} of the
q-d'Alembert
equation is a linear $\ast$-map
$$ \mbox{\boldmath $\varphi$} : span_{\cn}\{ {\bf e}, \bar{\bf e} \}\rightarrow
M_{q}$$
such that (\ref{dalembert}) holds. The space $span_{\cn}\{{\bf e}, \bar{\bf
e}\}$ is simply
the linear component of ${\cal P}(\cn)$. For both 0-forms and complex valued
solutions of the q-d'Alembert equation, there exists a conserved current:

\begin{prop}
Let $\mbox{\boldmath $\varphi$}$ be a solution of the q-d'Alembert equation.
Then the current 1-form $\bf j$
$$ {\bf j} := \overline{\mbox{\boldmath $\varphi$}}\wedge
d\mbox{\boldmath $\varphi$}-q^{-2}d\overline{\mbox{\boldmath $\varphi$}}\wedge
\mbox{\boldmath $\varphi$}$$
is conserved: $$\delta {\bf j} =0.$$
\end{prop}

{\bf Proof.}
Equation (\ref{xxx}), corollary \ref{leiwed},
 and the relation $q^{-2}{\bf R}^{-1cd}_{L\;\;\;\;ab}
g^{ba}=g^{cd}$  imply:
$$
\begin{array}{rcl}
\delta {\bf j}({\bf e}_{a}) &=&
\partial_{b}(\overline{\varphi}\partial_{a}\varphi
-q^{-2}(\partial_{a}\overline{\varphi})\varphi)H^{a}_{\{1\} cde}
{\cal A}^{\;bcde}_{\{4\} klmn}H^{klmn}_{\{4\}}\\
&=& \partial_{b}(\overline{\varphi}\partial_{a}\varphi
-q^{-2}(\partial_{a}\overline{\varphi})\varphi)H^{a}_{\{1\} cde}
H^{bcde}_{\{4\}} H_{\{0\} klmn}H^{klmn}_{\{4\}}\\
&=& \partial_{b}(\overline{\varphi}\partial_{a}\varphi
-q^{-2}(\partial_{a}\overline{\varphi})\varphi)g^{ba}\\
&=& ((\partial_{b} \overline{\varphi})(\partial_{a}\varphi)
-(\partial_{c}\overline{\varphi}) (\partial_{d} \varphi)
q^{-2}{\bf R}^{-1cd}_{L\;\;\;\;ab})g^{ba}\\
&=& 0
\end{array}
$$
Here we used repeatedly the relations (\ref{ho1}).
\hfill $\Box$

The interesting feature of this current is that  unlike in the classical case,
 it does {\it not}  vanish if $\mbox{\boldmath $\varphi$}$ is real.
If we assume that a non-trivial
$\mbox{\boldmath $\varphi$}$ is given by a central and real $\varphi \in
M_{q}$, then
$${\bf j} : {\bf e}_{a} \mapsto (1-q^{-2}) \varphi \partial_{a}\varphi, $$
vanishes only in the commutative case.

\subsection{Plane wave solutions}\label{plaw}

We shall now construct a family of plane wave solutions to the q-d'Alembert
equation.
For this purpose, we regard a copy of $V ({\bf R}_{M})$ as momentum space and
denote its generators by $p^{a}$. It is a
$\widetilde{\cal L}_{q}$-comodule algebra with coaction $p^{a}\mapsto
p^{b}\otimes S\lambda^{a}_{\;b}\varsigma^{-1}$, i.e. has the
$\varsigma$-scaling
property as appropriate for momenta.
The relations between the $p$'s are described in terms of ${\bf R}_{M}$, but
on the {\it q-deformed light cone} $P_{0}$ defined as the quotient of $V ({\bf
R}_{M})$
by the relation $g_{ab}p^{a}p^{b}=0,$
one also has:

\begin{lemma}\label{mshell}
There is an isomorphism
$$P_{0}\cong  V (q^{-2}{\bf R}_{L\;})/(g_{ab}p^{a}p^{b}=0). $$
Thus  on the quotient $P_{0}$, the generators also obey the
relations
$$p^{a}p^{b}=q^{-2}{\bf R}^{\;ab}_{L\;cd}p^{d}p^{c}.$$
\end{lemma}

{\bf Proof.}
Let $p=(a,b,c,d)$ be the vector of generators. The algebra
$V(q^{-2}{\bf R}_{L\;})$ has the same relations as $V( {\bf R}_{M})$ except for
$cb=q^{2}bc +(1-q^{2})dd$, which differs from the corresponding relation
$cb = bc - (1-q^{2}) ad - (1-q^{-2}) dd$. However, in the
quotients $ V (q^{-2}{\bf R}_{L\;})/(g_{ab}p^{a}p^{b}=0)$ and $P_{0}$,
 the generators obey $ad-q^{-2}cb=0$ and
 we can rewrite both relations as
$ad=bc-(1-q^{-2})dd$. \hfill $\Box$

The q-light cone is invariant under the
coaction by $\widetilde{\cal L}_{q}$, i.e.
the coaction $\beta$ by the q-Lorentz group on  $V( {\bf R}_{M})$
descends to a covariant coaction
$\beta: P_{0}\rightarrow P_{0}\otimes\widetilde{\cal L}_{q}$. Using $P_{0}$
as an `index set' we define a family of {\it  q-deformed plane waves}:
\begin{equation}
exp(i x.p):= \sum^{\infty}_{n=0} \frac{i^{n}}{[n]!} x_{1}\ldots x_{n}
p_{n}\ldots p_{1}
\end{equation}
as a formal power series in $M_{q}\underline{\otimes} P_{0}$,
where $[n]=1+q+\ldots + q^{n-1}$ and $[n]!=[1]\ldots [n]$. This exponential is
different
from the one proposed in \cite{DAMTP/93-3}, but is based on the same idea.

\begin{prop}\label{plawa}
The family of $P_{0}$-indexed complex valued plane waves
$$
\begin{array}{rcl}
\mbox{\boldmath $\varphi$}(p): span_{\cn}\{ {\bf e}, \bar{\bf e} \}
&\rightarrow & M_{q}\\
{\bf e} &\mapsto &exp( i x.p) \\
\bar{\bf e} &\mapsto &exp(- i x.p)
\end{array}
$$
are solutions of the q-d'Alembert equation.
\end{prop}

{\bf Proof.}
First note that  $exp(i x.p)$ transforms as a scalar under the coaction by
$\widetilde{\cal L}_{q}$, since the dilaton terms always cancel. Thus
$\mbox{\boldmath $\varphi$}(p)$ are $\widetilde{\cal L}_{q}$-comodule
morphisms,
and it is also obvious that they are $\ast$-maps.
It remains to show $\Box exp(ix.p)=0$:
$$\begin{array}{l}
\partial^{a} exp(ix.p)\\
= \sum_{n} i^{n}\frac{1}{[n]!}\partial^{a}x_{1}\ldots x_{n} p_{n}\ldots p_{1}\\
= \sum_{n} i^{n}\frac{1}{[n]!} \delta^{a}_{\;1}x_{2}\ldots x_{n}
(1+q^{-1}P{\bf R}^{-1}_{L\; 12}+\ldots +q^{-(n-1)} P{\bf R}^{-1}_{L\; 12}\ldots
P{\bf R}^{-1}_{L\; (n-1)n} ) p_{n}\ldots p_{1}\\
= \sum_{n} i^{n}\frac{1+q+\ldots+q^{n-1}}{[n]!}x_{2}\ldots x_{n}
 p_{n}\ldots p_{2} \;\; p^{a}\\
= exp(x.p)\;\; ip^{a}
\end{array}$$
 and thus
$\Box exp(ix.p) = 0.$
\hfill $\Box$

These plane wave type solutions exist {\it only} on the q-light cone, giving
further support
to our claim that wave equations on quantum Minkowski space should be massless.
In this special case of plane wave solutions, we can also show a stronger
statement
than proposition \ref{plawa}. In general, we
require $\mbox{\boldmath $\varphi$}$ only to be a comodule morphism on the
linear
component   of ${\cal P}(\cn)$, but the plane wave solutions   can be
extended  to  the whole algebra:

\begin{prop}
The linear maps $\mbox{\boldmath $\varphi$}(p)$ extend to $\ast$-algebra maps
$$ \mbox{\boldmath $\varphi$}(p) : {\cal P}(\cn) \rightarrow M_{q}, $$
which are also $\widetilde{\cal L}_{q}$  comodule morphisms.
\end{prop}

{\bf Proof.} It is sufficient to
show that $exp(-ix.p)$ and $exp(ix.p)$ commute.
Using the statistics relations $p_{1}x_{2}=x_{2}{\bf R}^{-1}_{L\; 12}p_{1}$
and lemma \ref{mshell}, which implies
$$
q{\bf R}^{-1}_{L\;  12}
p_{1}p_{2}=q^{-1}p_{2}p_{1},\;\;\;\;{\bf R}_{M 12}p_{2}p_{1}=
q^{2}{\bf R}^{-1}_{L\; 12}p_{2}p_{1},
$$
 we can show that the monomials commute:
$$\begin{array}{l}
(x_{1}\ldots x_{n} p_{n}\ldots p_{1})
(x_{1^{'}}\ldots x_{m^{'}}
p_{m^{'}}\ldots p_{1^{'}})\\
=x_{1}\ldots x_{n}x_{1^{'}}\ldots x_{m^{'}}
(q{\bf R}^{-1}_{L\;   nm^{'}}\ldots
q{\bf R}^{-1}_{L\; 1m^{'}})\ldots
(q{\bf R}^{-1}_{L\; n1^{'}}\ldots
q{\bf R}^{-1}_{L\; 11^{'}}) p_{n}\ldots p_{1}
p_{m^{'}}\ldots p_{1^{'}}\\
= q^{-nm}x_{1}\ldots x_{n}x_{1^{'}}\ldots x_{m^{'}}
 p_{m^{'}}\ldots p_{1^{'}} p_{n}\ldots p_{1}\\
= q^{-nm} x_{1^{'}}\ldots x_{m^{'}}x_{1}\ldots x_{n}
({\bf R}_{M\; 1m^{'}}\ldots {\bf R}_{M\; nm^{'}})\ldots
({\bf R}_{M\; 11^{'}}\ldots {\bf R}_{M\; n1^{'}})
 p_{m^{'}}\ldots p_{1^{'}} p_{n}\ldots p_{1}\\
=  x_{1^{'}}\ldots x_{m^{'}}x_{1}\ldots x_{n}
(q{\bf R}^{-1}_{L\; m^{'}1}\ldots q{\bf R}^{-1}_{L\; m^{'}n})
\ldots
(q{\bf R}^{-1}_{L\; 1^{'}1}\ldots q{\bf R}^{-1}_{L\; 1^{'}n})
  p_{m^{'}}\ldots p_{1^{'}} p_{n}\ldots p_{1}\\
= (x_{1^{'}}\ldots x_{m^{'}}
p_{m^{'}}\ldots p_{1^{'}})
(x_{1}\ldots x_{n} p_{n}\ldots p_{1})
\end{array}
$$
We used primed and unprimed indices to distinguish the two monomials.
\hfill $\Box$

\section{q-Vector field}

\subsection{The q-Maxwell  equation }

For q-Maxwell equations, we apply a similar strategy as for the q-d'Alembert
equation: we first give a more abstract definition in terms of
$\delta$ and $d$ and then show how this equation looks in terms of the maybe
more familiar braided differential operators $\partial$.

\begin{defi}
A solution of the {\bf q-Maxwell equation} is a 1-form $\bf A$
such that
\begin{equation}\label{maxwell}
\delta d {\bf A} =0.
\end{equation}
\end{defi}

Using the results from the preceding sections, we can rewrite this rather
abstract relation   to resemble the
classical equation $\partial^{\mu}\partial_{\mu}A_{\nu}-
\partial^{\mu}\partial_{\nu}A_{\mu}=\partial^{\mu}\partial_{[\mu}A_{\nu]}=0$:

\begin{prop}\label{MAX}
The equation (\ref{maxwell}) is equivalent to the
set of four equations
\begin{equation}\label{max}
\partial^{c} \partial_{[c} A_{z]} = 0,
\end{equation}
or alternatively
\begin{equation}\label{max2}
\Box A_{z} - \partial_{z}\partial^{c}A_{c} =0.
\end{equation}
Here $\partial$ denotes the braided differential operators on $M_{q}$, $\Box$
the
q-d'Alembert operator  and ` $[\;\;]$' the q-antisymmetriser.
\end{prop}

{\bf Proof.}
First verify by explicit calculation that
\begin{equation}\label{HO2}
H^{\;ab}_{\{2\} cd}H^{\;xcd}_{\{3\}\;\;\;\; z}=
\frac{(1+q^{2})^{2}}{q (2 (1+q^{2}+q^{4}))^{1/2}}g^{xc} {\cal A}^{\;ab}_{\{2\}
cz}.
\end{equation}
By virtue of this relation, we obtain
$$
\begin{array}{rcl}
0&=&\delta d {\bf A} ({\bf e}_{z})\\
&=& \partial_{x}\partial_{a}A_{a} {\cal A}^{\;ab}_{\{2\} cd}H^{\;cd}_{\{2\} ef}
{\cal A}^{\;xef}_{\{3\}klm}H^{\;klm}_{\{3\}\;\;\;  z}\\
&=& \partial_{x}\partial_{a}A_{b} H^{\;ab}_{\{2\} cd}{\cal A}^{\;cd}_{\{2\} ef}
{\cal A}^{\;xef}_{\{3\}klm}H^{\;klm}_{\{3\}\;\;\; z}\\
&=& \partial_{x}\partial_{a}A_{b} H^{\;ab}_{\{2\} ef}
{\cal A}^{\;xef}_{\{3\}klm}H^{\;klm}_{\{3\}\;\;\;  z}\\
&=& \partial_{x}\partial_{a}A_{b} H^{\;ab}_{\{2\} ef}
H^{\;xef}_{\{3\}\;\;\;\;  j} H^{\;j}_{\{1\} klm}H^{\;klm}_{\{3\}\;\;\;  z}\\
&=& \partial_{x}\partial_{a}A_{b} H^{\;ab}_{\{2\} ef}
H^{\;xef}_{\{3\}\;\;\; z}\\
&=& \partial_{x}\partial_{a}A_{b}g^{xc}{\cal A}^{\;ab}_{\{2\} cz}\\
&=&\partial^{c} \partial_{[c} A_{z]},
\end{array}
$$
where we used
 the definitions of $\delta$ and $d$, the relations (\ref{ho1}) and
(\ref{cancel}),
and finally (\ref{HO2}). This establishes the equivalence of (\ref{max}) and
(\ref{maxwell}). In order to prove (\ref{max2}) note that the generators of
$V({\bf R}_{M})$ obey
\begin{equation}\label{ler}
x_{a}x^{b}=x^{c}x_{d}{\bf R}^{\;\; bd}_{M\;ca}.
\end{equation}
Hence with (\ref{nohe}), we find:
$$
\begin{array}{rcl}
0&=&\partial^{c} \partial_{[c} A_{z]}\\
&=& \partial^{c} (\partial_{c}A_{z} - \partial_{i}A_{j}{\bf R}^{\;\;ji}_{M\;
cz})\\
&=& \partial^{c} \partial_{c}A_{z} -\partial^{c}\partial_{i}A_{j}{\bf
R}^{\;\;ji}_{M\; cz}\\
&=& \Box A_{z} - \partial_{z}\partial^{c}A_{c}
\end{array}
$$
\hfill $\Box$

In the form (\ref{max}), the  $\widetilde{\cal L}_{q}$-covariance of the
q-Minkowski equations can be easily
established. Again, a massive field equation, i.e. a {\it q-Proca equation}
$$ \partial^{c} \partial_{[c} A_{z]} = m^{2} A_{z}$$
would be $\widetilde{\cal L}_{q}$-covariant only if $m$ transformed as
$m\mapsto m\otimes\varsigma^{-1}$, and again we shall find q-deformed
plane wave solutions only on the q-light cone.

As in the undeformed case, solutions to the q-Maxwell equation have a {\it
gauge
freedom}.
If ${\bf A}$ is a solution of the  q-Maxwell equation and
$\mbox{\boldmath $\varphi$}$ a  0-form  then by virtue of theorem \ref{exclo}
the 1-form
${\bf A} + d\mbox{\boldmath $\varphi$}$
is also a solution.
Provided it is possible to solve the inhomogeneous  equation
$$ \mbox{\boldmath $\Delta$}\mbox{\boldmath $\varphi$} =-\delta {\bf A} ,$$
we can use this gauge freedom to arrange for ${\bf A}$ to satisfy
the {\it q-Lorentz gauge} condition
$$\delta {\bf A}=0 .$$
Using an argument similar to the proof of proposition \ref{obvi}, one can
show that the q-Lorentz gauge condition is satisfied iff
\begin{equation}\label{qlo}
\partial^{c}A_{c}=0 .
\end{equation}
Proposition \ref{MAX} implies that in this case $\bf A$ obeys
 $\mbox{\boldmath $\Delta$} {\bf A}=0$ or equivalently
$$\Box A_{z}=0.$$
As in the classical case, a field  ${\bf A}$ satisfying the q-Lorentz gauge
 has a {\it  residual gauge freedom }
$${\bf A} \mapsto {\bf A} + d\mbox{\boldmath $\varphi$},$$
where $\mbox{\boldmath $\varphi$}$ is a solution of the q-d'Alembert  equation.

The q-Maxwell equation also has a family of plane wave solutions. However, in
this case
the solutions are indexed by the q-momentum $p^{a}$ (the generators of the
q-light cone
$P_{0}$) and the `q-amplitude' $A_{z}$, which are the generators of a copy of
$M_{q}$.
We define the algebra $Y$ as the  quotient of $P_{0}\underline{\otimes}M_{q}$
by the relation
$$p^{c}\underline{\otimes} A_{c} =0.$$
This algebra $Y$ labels the plane wave solutions to the q-Maxwell equation:

\begin{prop}
Then the family of $Y$-indexed 1-forms
$$ {\bf A} : {\bf e}_{z} \mapsto exp(i x.p)\underline{\otimes} A_{z}$$
viewed as a formal power series in $M_{q}\underline{\otimes} Y$ are
solutions of the q-Maxwell equation and satisfy the
q-Lorentz gauge condition  $\delta {\bf A}=0$.
\end{prop}

{\bf Proof.}
Using the q-Maxwell equations in the form (\ref{max}), one finds:
$$
\begin{array}{rcl}
\partial^{c}\partial_{[c} exp(i x.p)\underline{\otimes} A_{z]}
&=& \partial^{c}(\partial_{c} exp(i x.p)\underline{\otimes} A_{z}
- \partial_{m} exp(i x.p)\underline{\otimes} A_{n}{\bf R}^{\;\;nm}_{M\;cz})\\
&=&  exp(i x.p)p^{c}p_{m}\underline{\otimes} A_{n}{\bf R}^{\;\;nm}_{M\;cz}\\
&=&  exp(i x.p)p_{z}p^{c}\underline{\otimes} A_{c}\\
&=& 0.
\end{array}
$$
Here we used (\ref{nohe}), proposition \ref{plawa} and relation (\ref{ler}).
These
solutions obviously satisfy the q-Lorentz gauge condition.
\hfill $\Box$

A solution $\bf A$ of the q-Maxwell equation (\ref{maxwell})
defines a 2-form ${\bf F} = d{\bf A}$, the {\it q-field strength tensor}  which
obeys
the two equations
\begin{equation}\label{maxf}
d{\bf F}=0,\;\;\;\; \delta {\bf F} =0
\end{equation}
Proposition \ref{MAX}
implies that the second relation is equivalent to
\begin{equation}\label{divf}
\partial^{c}F_{cd}=0.
\end{equation}
Furthermore, we find
$$
\begin{array}{rcl}
\Box F_{ab} &=& \Box \partial_{[a}A_{b]}\\
&=&\partial_{[a} \Box A_{b]}\\
&=&\partial_{[a} \partial_{b]} \partial^{c}A_{c}\\
&=&0,
\end{array}
$$
using the fact that $\Box$ is central in ${\cal D}$ and the q-Maxwell equation
for $\bf A$.
Since at present we do not have
a q-Poincar\'{e} lemma, we only know that q-Maxwell equations in the form
(\ref{maxf}) are implied by (\ref{maxwell}), but we cannot prove that they are
equivalent.
But nevertheless, we shall proceed by  investigating the q-Maxwell equation in
terms of
this q-field strength tensor.

\subsection{Elements of the $SL_{q}(2,\cn)$-spinor calculus}

In the next section  we shall give a $SL_{q}(2,\cn)$-spinor description of the
q-field
strength tensor
 ${\bf F}$ similar to the  the classical case.  For this purpose, we need some
elements
of the $SL_{q}(2,\cn)$-spinor calculus, some aspects of which were already
discussed in \cite{Schlieker/2/90}.
This case is very simple since the R-matrix (\ref{rmat}) is of Hecke-type, i.e.
obeys
\begin{equation}\label{Hecke}
0=(PR+q^{-1})(PR-q).
\end{equation}
This means  that one can take either $(PR-q)$ or $(PR^{-1}-q^{-1})$
as a  q-antisymmetriser for
$SL_{q}(2,\cn)$-spinors. Relation (\ref{Hecke}) ensures that
after a suitable normalisation these operators are
projectors. Furthermore, one does not have any problems with higher
q-antisymmetrisers, since they do not exist.
One could also define a q-antisymmetriser by first  identifying a
q-antisymmetric
$\varepsilon_{AB}$, similar to the procedure in section \ref{asy}, but this
approach gives
the same result.
The q-antisymmetric spinor
$\varepsilon_{AB}$ is simply the  $SL_{q}(2,\cn)$-spinor metric (\ref{spim}),
which  obeys
\begin{equation}\label{epsym}
q\varepsilon_{AB}=
-\varepsilon_{CD} R^{-1DC}_{\;\;\;\;\;AB}.
\end{equation}
One can   easily verify that then
$$A^{AB}_{\; CD}=\frac{1}{q+q^{-1}}\varepsilon^{BA}\varepsilon_{CD}=
\frac{1}{q+q^{-1}}(PR^{-1}-q^{-1})^{AB}_{\;CD}.$$
obeys  $A^{2}=A$ by virtue of  (\ref{Hecke}).
We  also define
define a {\it q-symmetriser}
$$S:= \frac{1}{2} (1-A) = \frac{1}{q^{-1}+q}(PR^{-1}+q),$$
and accordingly the {\it q-symmetrisation} `$(\;\;)$'  and {\it
q-antisymmetrisation}
`$[\;\;]$' of a multivalent
$q$-spinor $T_{\ldots AB\ldots}$ with two adjacent lower indices $A$ and $B$ as
$$
T_{\ldots (AB) \ldots}=
T_{\ldots CD\ldots}S^{CD}_{\;AB},\;\;\;\;
T_{\ldots [AB] \ldots}= T_{\ldots CD\ldots}A^{CD}_{\;AB}
$$and similarly for upper indices.
Due to (\ref{Hecke}), the
$q$-(anti)-symmetrisation of a $q$-spinor is  $q$-(anti)-symmetric:
\begin{equation}\label{symasym}
T_{\ldots \;[\;(AB)\;]\; \ldots}=0,\;\;\;\;\;T_{\ldots \;(\;[AB]\;)
\;\ldots}=0,
\end{equation}
and we also have a decomposition
\begin{equation}\label{decomp}
T_{\ldots AB \ldots}= T_{\ldots (AB) \ldots}+T_{\ldots [AB] \ldots}.
\end{equation}
The q-(anti)-symmetrisation is  also $SL_{q}(2,\cn)$-covariant, i.e. both
operations
`$(\;\;)$' and `$[\;\;]$' commute with the coaction
by $SL_{q}(2,\cn)$.
Furthermore, if  $T_{\ldots CD\ldots}$ is a multivalent $q$-spinor, then
\begin{equation}\label{cool}
  T_{\ldots \; [CD]\; \ldots } =  \frac{1}{q^{-1}+q}\;
\varepsilon_{CD}\;T_{\;\ldots B \;\;\;
\ldots}^{\;\;\;\;\;\; B}.
\end{equation}
In this formula we do not violate the index notation by writing
$\varepsilon_{CD}$
on the left since $C$ and $D$
are adjacent indices and the generators of $SL_{q}(2,\cn)$ preserve the spinor
metric.

\subsection{q-Spinor analysis of the q-field strength tensor}

We now apply the results from the last section to the field strength tensor
$\bf F$, or more generally, to any q-antisymmetric tensor $F_{ab}\in M_{q}$.
Similar to the classical case, such a tensor decomposes into
$SL_{q}(2,\cn)$-spinors:

\begin{prop}\label{decom}
Let $F_{ab} \in M_{q}$ be an q-antisymmetric  $\widetilde{\cal L}_{q}$-tensor
in the
sense of (\ref{anti}). Then
$${f^{AB}}_{\;A^{'}B^{'}} := F_{AI^{'} I B^{'}}
R^{I^{'}  B}_{\; A^{'}I}$$
admits a decomposition
$${f^{AB}}_{\;A^{'}B^{'}}  = \phi^{AB}\varepsilon_{A^{'}B^{'}}+
\varepsilon^{AB}\psi_{A^{'}B^{'}},$$
where $ \phi^{AB}$ and $\psi_{A^{'}B^{'}}$ are q-symmetric
$SL_{q}(2,\cn)$-spinors.
\end{prop}

{\bf Proof.}
Since the tensor $F_{ab} $ is q-antisymmetric, ${f^{AB}}_{\;A^{'}B^{'}} $
obeys:
\begin{equation}\label{pro}
\begin{array}{rcl}
{f^{AB}}_{\;A^{'}B ^{'}}
&=& F_{AI^{'} I B^{'}}R^{I^{'}  B}_{\; A^{'}I}\\
&=& -F_{C C^{'}D D^{'}}{\bf R}^{DD^{'}CC^{'}}_{L\;AI^{'}  IB^{'}}  R^{I^{'}
B}_{\; A^{'}I}\\
 &=& -  F_{C C^{'}D D^{'}}
R^{C^{'} K}_{\;L  D}R^{D^{'} L }_{\;M B^{'}}R^{A N}_{\;KC}
\widetilde{R}^{MI}_{\;I^{'}N}  R^{I^{'}  B}_{\; A^{'}I}\\
 &=& -  {f^{CD}}_{\;C^{'}D^{'}}
R^{D^{'}C^{'}}_{\;A^{'}B^{'}}R^{AB}_{\;DC}
\end{array}
\end{equation}
Due to relation (\ref{decomp}), we also have
$${f^{AB}}_{A^{'}B^{'}}= {f^{(AB)}}_{(A^{'}B^{'})}+
{f^{[AB]}}_{(A^{'}B^{'})} + {f^{(AB)}}_{[A^{'}B^{'}]}+
{f^{[AB]}}_{[A^{'}B^{'}]}.$$
This implies with (\ref{pro}), (\ref{symasym}) and (\ref{Hecke}):
$$
\begin{array}{rcl}
{f^{AB}}_{A^{'}B^{'}}&=& -  {f^{CD}}_{\;C^{'}D^{'}}
R^{D^{'}C^{'}}_{\;A^{'}B^{'}}R^{AB}_{\;DC} \\
&=& -q^{2} {f^{(AB)}}_{(A^{'}B^{'})}+
{f^{[AB]}}_{(A^{'}B^{'})} + {f^{(AB)}}_{[A^{'}B^{'}]}-q^{-2}
{f^{[AB]}}_{[A^{'}B^{'}]},
\end{array}
$$
and also
$$
\begin{array}{rcl}
{f^{AB}}_{A^{'}B^{'}}&=& -  {f^{CD}}_{\;C^{'}D^{'}}
R^{-1  C^{'}D^{'}}_{\; \;\;\;\;B^{'}A^{'}}R^{-1 BA }_{\;\;\;\;\;\; CD } \\
&=& -q^{-2} {f^{(AB)}}_{(A^{'}B^{'})}+
{f^{[AB]}}_{(A^{'}B^{'})} + {f^{(AB)}}_{[A^{'}B^{'}]}-q^{2}
{f^{[AB]}}_{[A^{'}B^{'}]},
\end{array}
$$
and therefore $0={f^{(AB)}}_{(A^{'}B^{'})}+{f^{[AB]}}_{[A^{'}B^{'}]}$.
With
relation (\ref{cool}), it follows
$${f^{AB}}_{\;A^{'}B^{'}}  = \phi^{AB}\varepsilon_{A^{'}B^{'}}+
\varepsilon^{AB}\psi_{A^{'}B^{'}},$$
where
$$\phi^{AB} =  f^{(AB)\;\;\;C}_{\;\;\;\;\;\;\;\;\;C},\;\;\;\;\;\;
\psi_{A^{'}B^{'}} =  f^{\;\,\;C}_{C \;\;\;(A^{'}B^{'})}$$
are q-symmetric $SL_{q}(2,\cn)$-spinors.
\hfill $\Box$

In the case of a real tensor, the two components $\phi$ and $\psi$ are not
independent,
but are related by the star structure on $M_{q}$.
Recall that the conjugate of a tensor  $T_{a\ldots d}\in M_{q}$ is given by
$ \overline{T}_{a\ldots d} = T_{\bar{d}\ldots\bar{a}}$. Such a tensor is called
{\it real}
if $T=\overline{T}$.

\begin{prop}\label{recase}
A  q-antisymmetric tensor $F_{ab}$ is real iff
$$\psi_{DC} = -\bar{\phi}^{CD}.$$
Thus a real tensor can be written as
$${f^{AB}}_{\;A^{'}B^{'}}  = \phi^{AB}\varepsilon_{A^{'}B^{'}}+
\varepsilon_{BA}\bar{\phi}^{B^{'}A^{'}}$$
in terms of the q-symmetric $SL_{q}(2,\cn)$-spinor $\phi^{AB}$.
\end{prop}

{\bf Proof.}
For the proof, we need the symmetry property
\begin{equation}\label{rsym}
R^{AB}_{\;CD}=R^{DC}_{\;BA}
\end{equation}
of the $SU_{q}(2)$ R-matrix, which can be verified by inspection of
(\ref{rmat}).
A similar relation holds for $R^{-1}$ and $\widetilde{R}$.
Hence if $F_{ab}$ is real then
$$
\begin{array}{rcl}
{\bar{f}^{AB}}_{\;A^{'}B^{'}} &=& \overline{F}_{A I^{'}I B^{'}}
R^{I^{'} B}_{\; A^{'} I}\\
&=& F_{B^{'} II^{'} A}
R^{I^{'} B}_{\; A^{'}I}\\
&=& F_{B^{'}  II^{'} A}
R^{A^{'}I }_{\; BI^{'}}\\
&=&{f^{B^{'}A^{'}}}_{\;BA} .
\end{array}
$$
In components, this means that
$$ \bar{\phi}^{AB}\varepsilon_{A^{'}B^{'}}+
\varepsilon^{AB}\bar{\psi}_{A^{'}B^{'}}=
 \phi^{B^{'}A^{'}}\varepsilon_{BA}+
\varepsilon^{B^{'}A^{'}}\psi_{BA}.$$
Due to the q-symmetry of $\phi$ and $\psi$ and the q-antisymmetry of
$\varepsilon_{AB}$,
multiplication of this equation by $qR^{CD}_{\;BA}$ yields by virtue of
(\ref{symasym}):
$$ q^{2}\bar{\phi}^{CD}\varepsilon_{A^{'}B^{'}}-
\varepsilon^{CD}\bar{\psi}_{A^{'}B^{'}}=
-\phi^{B^{'}A^{'}}\varepsilon_{DC}+
q^{2}\varepsilon^{B^{'}A^{'}}\psi_{DC},$$
again using the relation (\ref{rsym}). Thus
$$\bar{\phi}^{CD}\varepsilon_{A^{'}B^{'}}=\varepsilon^{B^{'}A^{'}}\psi_{DC},$$
which implies the proposition, since
$\varepsilon^{B^{'}A^{'}}=-\varepsilon_{A^{'}B^{'}}$.
\hfill $\Box$

Classically, this decomposition of the field strength tensor into spinors
coincides with
the decomposition into its self-dual and anti-self-dual part.
The same result holds in the non-commutative case.
By virtue of proposition \ref{hodge2}, any two-form $\bf F$ on quantum
Minkowski space can be decomposed uniquely as
$${\bf F} = {\bf F}^{+}+{\bf F}^{-},$$
where  $ {\bf F}^{+} = \frac{1}{2}( {\bf F}+\,  ^{\ast}{\bf F})$
and $ {\bf F}^{-} = \frac{1}{2}( {\bf F}-\,  ^{\ast}{\bf F})$  are  self-dual
and anti-self-dual, i.e. obey
 $^{\ast}{\bf F}^{\pm}=\pm {\bf F}^{\pm}$.
The q-Maxwell equation  (\ref{maxf}) are then equivalent to either
$$ d{\bf F}^{+} = 0 ,\;\;\;\;\; d{\bf F}^{-} = 0,$$
or the two equations
\begin{equation}\label{maspi}
 \delta{\bf F}^{+} = 0 ,\;\;\;\;\; \delta{\bf F}^{-} = 0.
\end{equation}

\begin{prop}\label{selfanti}
Let $F_{AB}$ be a q-antisymmetric tensor. Then
$${f^{+\;AB}}_{\;\;\;A^{'}B^{'}}  = \phi^{AB}\varepsilon_{A^{'}B^{'}},\;\;\;\;
{f^{-\;AB}}_{\;\;\;A^{'}B^{'}}  = \varepsilon^{AB}\psi_{A^{'}B^{'}}$$
are the self-dual and anti-self-dual parts of $f$.
\end{prop}

{\bf Proof.}
It suffices to show that $f^{\pm}$ are selfdual and antiselfdual, respectively.
On the tensor ${f^{AB}}_{\;\;A^{'}B^{'}}$, the q-Hodge star operation is
implemented
by the matrix
$$    U^{\;ABA^{'}B^{'}}_{\{2\} CDC^{'}D^{'}} :=
\widetilde{R}^{A^{'}I}_{\;I^{'}B}H^{\;\;AI^{'}IB^{'}}_{\{2\} CJ^{'}JD^{'}}
R^{J^{'}D}_{\;C^{'}J}.$$
By explicit calculation, one verifies that this operator satisfies the
relations
$$
\begin{array}{rcl}
S^{AB}_{\;EF} \varepsilon_{A^{'}B^{'}}  U^{\;ABA^{'}B^{'}}_{\{2\} CDC^{'}D^{'}}
 &=& S^{CD}_{\;EF}
\varepsilon_{C^{'}D^{'}},\\
\\
\varepsilon^{AB}
S^{E^{'}F^{'}}_{\;A^{'}B^{'}}
 U^{\;ABA^{'}B^{'}}_{\{2\} CDC^{'}D^{'}}  &=&
\varepsilon^{CD}
S^{E^{'}F^{'}}_{\;C^{'}D^{'}}.
\end{array}
$$
Since $\phi^{AB}$ and $\psi_{A^{'}B^{'}}$ are q-symmetric and thus eigenvectors
of the
q-symmetriser $S$, this implies that
$\phi^{EF}\varepsilon_{A^{'}B^{'}}$ and $ \varepsilon^{AB}\psi_{A^{'}B^{'}}$
are
self-dual and anti-self-dual, respectively.
 \hfill $\Box$

If we are looking for real solutions of the q-Maxwell equations (\ref{maxf}),
it is thus
sufficient to solve one of the two equations in (\ref{maspi}).  In terms of the
$SL_{q}(2,\cn)$-spinor  $\psi$ this means:

\begin{cor}
For real $\bf F$, the q-Maxwell equation $\delta {\bf F}^{-}$ is equivalent to
either
\begin{equation}\label{spimax}
\nabla^{BI^{'}}\psi_{I^{'}B^{'}}=0
\end{equation}
or the conjugate equation
$$\phi^{B^{'}I^{'}} \overline{\nabla}_{I^{'}B}=0$$
where $\nabla^{CC^{'}} := \widetilde{R}^{C^{'}C}_{\;A^{'}A}\partial^{AA^{'}},$
and its
conjugate
$\overline{\nabla}_{C^{'}C}:=
\partial_{A^{'}A}\widetilde{R}^{AA^{'}}_{\;CC^{'}}$
acts from the right.
\end{cor}

{\bf Proof.}
Proposition \ref{selfanti} implies with (\ref{divf})
$$
\begin{array}{rcl}
0&=& \partial^{a} F^{-}_{ab}\\
&=&\widetilde{R}^{I^{'}B}_{\;A^{'}I}
{\partial_{A}}^{A^{'}}\varepsilon^{AI}\psi_{I^{'}B^{'}}\\
&=& \widetilde{R}^{I^{'}B}_{\;A^{'}A}\partial^{AA^{'}} \psi_{I^{'}B^{'}}\\
&=& \nabla^{BI^{'}}\psi_{I^{'}B^{'}}.
\end{array}
$$
Since the tensor $F_{ab}$ is real, proposition \ref{recase}
 implies for the conjugate equation:
$$
\begin{array}{rcl}
0&=&
\bar{\psi}_{I^{'}B^{'}}{\partial_{A^{'}}}^{A}
\varepsilon^{AI}\widetilde{R}^{I^{'}B}_{\;A^{'}I}\\
&=&\phi^{B^{'}I^{'}}{\partial_{A^{'}}}^{A}
\varepsilon_{IA}\widetilde{R}^{I^{'}B}_{\;A^{'}I}\\
&=&\phi^{B^{'}I^{'}}\partial_{A^{'}A}\widetilde{R}^{AA^{'}}_{\;BI^{'}}\\
&=&\phi^{B^{'}I^{'}} \overline{\nabla}_{I^{'}B},
\end{array}
$$
We used relation (\ref{rsym}) and $\varepsilon_{AB}=-\varepsilon^{BA}$.
\hfill $\Box$

Equation (\ref{spimax}) is also quite easy to generalise to an arbitrary
q-spinor field. We call
a totally q-symmetric spinor $\psi_{A_{1} \ldots A_{n}} $ satisfying
$$\nabla^{B A_{1}}\psi_{A_{1} \ldots A_{n} } = 0$$
a {\it massless q-spinor field} of spin $\frac{1}{2}n$.
Thus in particular,  we fine the {\it q-Weyl equation}
$$ \nabla^{B A}\psi_{A}=0.$$
Details of the general case will be discussed elsewhere.


\begin{thebibliography}{10}

\bibitem{Schlieker/6/91}
U.~Carow-Watamura, M.~Schlieker, M.~Scholl, and S.~Watamura.
\newblock A quantum {L}orentz group.
\newblock {\em Int. J. Mod. Phys.}, A(17):3081--3108, 1991.

\bibitem{Schlieker/3/90}
U.~Carow-Watamura, M.~Schlieker, W.~Scholl, and S.~Watamura.
\newblock Tensor representations of the quantum group {$SL_{q} (2,C)$} and
  quantum {M}inkowski space.
\newblock {\em Z. Physik}, C(48):159--165, 1990.

\bibitem{DAMTP/94-7}
A.~Kempf and S.~Majid.
\newblock Algebraic q-integration and {F}ourier theory on quantum and braided
  spaces.
\newblock preprint DAMTP/94-7, January 1994.

\bibitem{Majid/12/91}
S.~Majid.
\newblock Examples of braided groups and braided matrices.
\newblock {\em J. Math. Phys.}, 32:3246--3253, 1991.

\bibitem{Majid/6/92}
S.~Majid.
\newblock Beyond supersymmetry and quantum symmetry.
\newblock In M.-L. Ge and H.~J. {de Vega}, editors, {\em Proc. 5th Nankai
  Workshop, Tianjin, China}. World Scientific, June 1992.

\bibitem{DAMTP/92-65}
S.~Majid.
\newblock Braided momentum in the $q$-{P}oincar\'{e} group.
\newblock {\em J. Math. Phys}, 34:2045--2058, 1993.

\bibitem{DAMTP/93-3}
S.~Majid.
\newblock Free braided differential calculus, braided binomial theorem, and the
  braided exponential map.
\newblock {\em J. Math. Phys.}, 34(10):4843--4856, October 1993.

\bibitem{DAMTP/92-12}
S.~Majid.
\newblock Quantum and braided linear algebra.
\newblock {\em J. Math. Phys.}, 34:1176--1196, 1993.

\bibitem{DAMTP/92-48}
S.~Majid.
\newblock The quantum double as quantum mechanics.
\newblock {\em J. Geom. Phys.}, 13:169--202, 1994.

\bibitem{DAMTP/93-68}
S.~Majid and U.~Meyer.
\newblock Braided matrix structure of $q$-{M}inkowski space and
  $q$-{P}oincar\'{e} group.
\newblock preprint DAMTP/93-68 (to appear in {\it Z. Physik C}), December 1993.

\bibitem{DAMTP/93-45}
U.~Meyer.
\newblock $q$-{L}orentz group and braided coaddition on $q$-{M}inkowski space.
\newblock preprint DAMTP/93-45, July 1993.

\bibitem{Wess/92}
O.~Ogievetsky, W.~B. Schmidke, J.~Wess, and B.~Zumino.
\newblock $q$-{D}eformed {P}oincar\'{e} algebra.
\newblock {\em Comm. Math. Phys.}, 150:495--518, 1992.

\bibitem{Schlieker/2/90}
M.~Schlieker and M.~Scholl.
\newblock Spinor calculus for quantum groups.
\newblock {\em Z. Physik}, C(47):625--628, 1990.

\bibitem{Schlieker/5/91}
M.~Schlieker, W.~Weich, and R.~Weixler.
\newblock Inhomogeneous quantum groups.
\newblock {\em Z. Phys. C}, 53:79--82, 1992.

\bibitem{WessZumino/90/1}
J.~Wess and B.~Zumino.
\newblock Covariant differential calculus on the quantum hyperplane.
\newblock In {\em Recent Advances in Field Theory}, pages 302--312. Nucl. Phys.
  B (Proc. Suppl.) 18B, 1990.

\end{thebibliography}
\end{document}